# RRhGe (R=Tb, Dy, Er, Tm): An experimental and theoretical study

Sachin Gupta,[1] K. G. Suresh,[1,*] A. K. Nigam[2], A. V. Lukoyanov[3,4]

[1]Department of Physics, Indian Institute of Technology Bombay, Mumbai-400076, India

[2]Tata Institute of Fundamental Research, Homi Bhabha Road, Mumbai-400005, India

[3]Institute of Metal Physics, Russian Academy of Sciences, Ural Branch, Yekaterinburg - 620990, Russia

[4]Ural Federal University, Yekaterinburg - 620002, Russia

## Abstract

RRhGe (R=Tb, Dy, Er, Tm) compounds have been studied by a number of experimental probes and theoretical *ab initio* calculations. These compounds show very interesting magnetic and electrical properties. All the compounds are antiferromagnetic with some of them showing spin-reorientation transition at low temperatures. The magnetocaloric effect (MCE) estimated from magnetization data shows very good value in all these compounds. The electrical resistivity shows metallic behavior of these compounds. MR shows negative sign near ordering temperatures and positive at low temperatures. The electronic structure calculations accounting for electronic correlations in the 4f rare-earth shell reveal the closeness of the antiferromagnetic ground state and other types of magnetic orderings in the rare-earth sublattice.

*Corresponding author email: suresh@phy.iitb.ac.in

# 1. Introduction

The intriguing magnetic and electrical properties of rare earth based intermetallic compounds continue to draw the attention of scientific community. Some families of compounds of this series have been found to be promising for certain applications, in addition to their interesting fundamental phenomena. RTX (R= rare earth, T=3d/4d/5d elements, X=p-block elements) family of compounds is one of the above mentioned series of compounds. An exhaustive review on magnetic and related properties of RTX compounds has been published recently [1]. One can see that compounds of this series show very interesting magnetic and electrical properties in a wide range of temperatures, depending on their crystal structures. Some compounds of this series show large magnetocaloric effect (MCE) and magneto-resistance (MR). Here we report experimental and theoretical investigations on RRhGe system, which is a member of the RTX series. The structural and basic magnetic properties of some compounds of RRhGe series were reported previously [2-6]. The compounds of this series were reported to have TiNiSi type orthorhombic crystal structure with space group *Pnma*. All the atoms (R=Ce-Tm, Rh and Ge) in these compounds occupy 4c site. It has also been found that all compounds of this series show antiferromagnetic ordering at low temperatures. CeRhGe and NdRhGe are antiferromagnetic with collinear magnetic structure with moments pointing along *b*-axis and having Néel temperature ($T_N$) of 10 and 14 K, respectively [4]. Szytuła et al. [5] reported that TbRhGe has a sine-modulated spin wave magnetic structure with a propagation vector k= (0, 0.45, 0.11) at 1.7 K, which transforms into a magnetic structure with a wave vector k= (0, 0.44, 0) at 18 K. According to this report, TbRhGe shows antiferromagnetic ordering below 23 K. HoRhGe and ErRhGe show collinear magnetic structure with a propagation vectors k= (1/2, 0, 1/2) and (0, 1/2, 0), respectively [6]. The magnetic structure changes to incommensurate sine-modulated structure at 5 K for ErRhGe [6]. The magnetic properties of DyRhGe and TmRhGe are not reported yet. GdRhGe [7] has two successive magnetic phase transitions ($T_1$=32 K and $T_2$= 24 K), while HoRhGe [8] shows the antiferromagnetic transition at 5.5 K. Large non-saturating positive MR was observed in GdRhGe at low temperatures, which is associated with complex magnetic structure below $T_1$, while HoRhGe shows negative MR near the antiferromagnetic transition and positive MR at 2 K. The electronic band structure calculations and optical properties reveal that Rh has negligible magnetic moment in HoRhGe compound [8].

In this work, we have carried out a systematic study of RRhGe (R=Tb, Dy, Er, Tm) compounds by a number of experiments along with electronic band structure calculations.

## 2. Experimental and computational details

All polycrystalline samples were prepared by arc melting of the constituent elements taken in their stoichiometric amounts in a water-cooled copper hearth under a continuous flow of purified argon gas with a titanium button used as the oxygen getter. The purity of starting elements was at least 99.9 at %. As-cast samples were sealed in evacuated quartz tube and annealed for 8 days at 800 ℃ followed by furnace cooling to improve the homogeneity. The phase purity of the sample were checked by the analysis of room temperature powder x-ray diffraction (XRD) pattern collected from X'Pert PRO diffractometer using Cu K$\alpha$ ($\lambda$=1.54 Å) radiation. DC magnetization, M (T, H) measurements were performed using Quantum Design, Physical Property Measurement System (PPMS). Heat capacity measurements were also carried out on PPMS using thermal relaxation method. The electrical resistivity was measured using the standard four-probe technique, applying an excitation current of 150 mA. The ac susceptibility (ACS) measurements were performed on Quantum Design, SQUID VSM under zero-field-cooled (ZFC) condition at various frequencies (f) and constant amplitude of ac field ($H_{ac}$).

Theoretical *ab initio* calculations of the electronic structure of RRhGe (R=Tb, Dy, Er, Tm) were performed with the LSDA+*U* method [9], combining the local spin density approximation (LSDA) and Hubbard *U* correction for strong electronic correlations of 4f electrons in rare-earth ions. The LSDA+*U* method was employed in both TB-LMTO-ASA [10] and PW-SCF Quantum-Espresso [11] packages. In TB-LMTO the orbital basis set included the muffin-tin orbitals corresponding to R (6s, 6p, 5d, 4f) states, Rh (5s, 5p, 4d, 4f) states, and Ge (4s, 4p, 4d) states. The muffin-tin orbitals radii were R(R) = 3.6 a.u. and R(Rh,Ge) = 2.7 a.u. and spin-orbit coupling was not included in these calculations. The values of the Coulomb parameter: $U$ = 5.7 (Tb); 5.8 (Dy); 6.5 (Er); 6.2 (Tm) eV, and exchange Hund parameter $J_H$ = 0.7 eV for all the compounds were calculated for 4f shell in separate constrained LDA calculations [12] for the experimental crystal structure parameters. The computed *U* and $J_H$ values were used in the LSDA+*U* calculations to account for electronic correlations in the 4f shell of rare-earth ions.

## 3. Results and discussion

The Rietveld refinement of XRD patterns shows that all the compounds are single phase with no detectable impurities. All the compounds are found to crystallize in the TiNiSi type orthorhombic structure with space group *Pnma* (No. 62). The lattice parameters obtained from least square fit are close to the reported values [3]. The XRD pattern along with Rietveld refinement for TbRhGe, as the representative compound, is shown in Fig. 1.

The dc magnetic susceptibility (DCS) for these compounds was measured in a constant dc field (500 Oe) in the temperature range 1.8-300 K and is shown in Fig. 2 along with the Curie-Weiss fit. The cusp-like nature at low temperatures of the DCS data of all the compounds shows that antiferromagnetic ordering is predominant in these compounds. Except TmRhGe, all the compounds of this series show an upturn at low temperatures. The upturn arises due to spin reorientation, which was also suggested by reported neutron diffraction data [5, 6]. The spin reorientation transition temperature is denoted as $T_t$. The inverse susceptibility data were fitted by the Curie-Weiss law (solid line in Fig. 2), $\chi^{-1}=(T-\theta_p)/C_m$ (where, $C_m$ is the molar Curie constant and $\theta_p$ is the paramagnetic Curie temperature) from which effective magnetic moment ($\mu_{eff.}$) and $\theta_p$ have been calculated (see Table I). One can see from Table I that the value of observed $\mu_{eff.}$ is close to the theoretical value of the respective free rare earth ion. From Fig. 2(a), it can be noted that at low fields, there is a small difference between ZFC and field cooled (FC) plots in TbRhGe, which disappears at 20 kOe. A similar behavior has been seen in some rare earth intermetallics as well [13, 14], which arises due to the domain wall pinning effect. The inset in Fig 2(b) shows derivative of magnetic susceptibility, which has its maximum at 20.3 K and minimum at 4.4 K, corresponding to antiferromagnetic and spin-reorientation transitions. The magnetic ordering temperature ($T_N$) estimated from $d\chi/dT$ vs. T plot is given in Table I. All the compounds have negative $\theta_p$, which confirms the antiferromagnetic ordering. It has been observed that TbRhGe has the highest $T_N$. It is of interest to note that TbRhSn has the highest ordering temperature in RRhSn series, which is quite identical to the RRhGe series [15].

**Table I.** Values of magnetic ordering temperature ($T_N$), paramagnetic Curie temperature ($\theta_p$), observed effective magnetic moment ($\mu_{eff}$), calculated free ion moment ($g\sqrt{J(J+1)}$), effective magnetic moment from the LSDA+*U* calculations ($\mu_{eff\,LSDA+U}$), critical field ($H_c$), -$\Delta S_M$ and $T_{ad}$

for field 50 kOe of RRhGe compounds.

| Compound | $T_N$ (K) | $\theta_p$ (K) | $\mu_{eff}$ ($\mu_B/R^{3+}$) | $g\sqrt{J(J+1)}$ ($\mu_B/R^{3+}$) | $\mu_{eff}$ LSDA+U ($\mu_B/R^{3+}$) | $H_C$ (kOe) | | $-\Delta S_M$ (J/kg K) | $\Delta T_{ad}$ (K) |
|---|---|---|---|---|---|---|---|---|---|
| | | | | | | $H_{C1}$ | $H_{C2}$ | | |
| TbRhGe | 24.3 | -18.9 | 10 | 9.72 | 9.6 | 48.6 | 54 | 1.9 | 1.3 |
| DyRhGe | 20.3 | -3 | 10.9 | 10.63 | 10.6 | 31.5 | 64.1 | 3.3 | 1.1 |
| ErRhGe | 10.2 | -9.6 | 9.7 | 9.59 | 9.5 | 14 | - | 7 | 6 |
| TmRhGe | 6.8 | -4.4 | 7.6 | 7.57 | 7.6 | 15 | - | 6.6 | 9.1 |

To get the exact nature of the magnetic phase transition and information about the magnetization dynamics of these compounds, we have performed the ac susceptibility measurements. The measurements were done in ZFC condition and in small amplitude of $H_{ac}$, which does not disturb the magnetic state of the system. We have performed ACS measurement for all the compounds at different frequencies (f=25-625 Hz) and with constant amplitude of ac field ($H_{ac}$= 1Oe) in the temperature range of 1.8- 50 K, as shown in Fig 3. One can see that there is single peak followed by an upturn at low temperatures in the real component of ACS, $\chi'$ (T). The peak in all the compounds coincides with the antiferromagnetic transition as determined in the DCS measurements. It is worth noting that the absence of the peak in $\chi''$ (T) (which is generally expected for collinear antiferromagnets) corresponding to the ordering temperature is another confirmation of antiferromagnetic nature of these phase transitions [16]. DyRhGe and ErRhGe show different behavior at low temperatures in $\chi''$ (T), which shows small peaks below their ordering temperature. Such maxima reflect energy losses in the magnetically ordered regime and arise due to the extra or intra-domain effects (i.e., domain wall motion or the domain rotation) seen in ferro, ferri, spin glass or canted systems [16, 17]. Maximum in $\chi''$ (T) has not been observed in TbRhGe, possibly due to the weak canting of the moments.

Fig. 4 shows the magnetization isotherms at selected temperatures below and above the ordering temperature obtained for fields up to 70 kOe for all the compounds. For the sake of clarity, the magnetization data only at lower temperatures have been shown with decreasing and increasing fields. As can be seen from these plots, initially the magnetization increases linearly with field in all the compounds, revealing the antiferromagnetic behavior of the compounds, while at higher fields they show slope change in magnetization due to the field induced metamagnetic transition. The critical magnetic field ($H_C$) for the metamagnetic transition has been calculated from dM/dH curve and the values estimated at lowest temperature are given in Table I. It can be seen from the table that DyRhGe and TbRhGe have two-step metamagnetic transitions with critical fields $H_{C1}$ and $H_{C2}$. The value of $H_{C1}$ in DyRhGe decreases with increase in temperature, while $H_{C2}$ increases with increase in temperature [see inset of Fig 4(b)]. In TbRhGe variation in $H_C$ is negligible with temperature. All the compounds show non-saturation trend even at 70 kOe, revealing that metamagnetic transition is not able to suppress the antiferromagnetic phase completely.

The heat capacity measurements were performed with and without field in the temperature range of 2- 100 K. The C vs. T plots for all the compounds are shown in Fig. 5. The λ-like peak on the onset of magnetic ordering in these compounds hints at the second order magnetic phase transition and is supported by non-hysteresis magnetization data. In TbRhGe, a small hump has been seen in the low temperature heat capacity data [see the inset of Fig 5(a)], which may be due to the reorientation of the moments, as also revealed by the DCS and ACS data. DyRhGe shows an upturn in C/T vs. T plot near 4 K, due to the second transition at 4 K seen in DCS and ACS measurements. Similarly a small kink is observed in ErRhGe [see inset of Fig. 5(c)]. These anomalies in heat capacity data are due to canting/reorientation of spins at low temperatures in these compounds. There is no anomaly in the C vs. T data of TmRhGe, reflecting the same behavior as shown by magnetization data. On application of field the peaks in heat capacity data get reduced and shift towards low temperatures revealing the antiferromagnetic behaviors of these compounds.

$\Delta S_M$ from M-H-T data has been estimated from the Maxwell's relation [18] given as

$$\Delta S_M = \int_0^H \left[ \frac{\partial M}{\partial T} \right]_H dH \qquad (1)$$

The information about the temperature and field dependencies of MCE is not only useful in applications but can also in understanding the underlying magnetism aspects. One can see from Fig. 6 that MCE is negative at low temperatures in TbRhGe, DyRhGe and ErRhGe (as expected for antiferromagnetic materials.), which manifests the magnetization and heat capacity data. At temperatures close to the antiferromagnetic transition temperature, the MCE is positive, which is generally observed for ferromagnetic materials. Thus, the application of field suppresses the antiferromagnetic interactions and gives rise to positive MCE. It is clear from MCE plots that the antiferromagnetic coupling is weak in the case of ErRhGe and TmRhGe compounds.

In order to get further understanding about the magnetic state of this series, electrical resistivity and MR measurements were performed. The electrical resistivity was measured in zero field as well as in presence various fields in the temperature range of 2-300 K. The magnetoresistance (MR) has been calculated from the field dependence of the resistivity data and is given by the formula as

$$MR = \left(\frac{\rho(T,H) - \rho(T,0)}{\rho(T,0)}\right) \times 100 \qquad (2)$$

Fig. 7 shows the temperature dependence of electrical resistivity for the RRhGe compounds in zero field. All the resistivity curves show a linear behavior in the high temperature range, indicating the metallic nature of the compounds. The slope change in resistivity data has been observed at the onset of the antiferromagnetic ordering. Fig. 8 shows the variation of MR % with the field. In TbRhGe, in paramagnetic regime, the magnitude of MR increases with decrease in temperature as one can see from Fig. 8(a). At 40 K, the value of MR is 2%, which increases to 4% at 30 K. Below the ordering temperature, the magnitude of MR is positive, increasing with increase in field and giving rise to a maximum value of 20% at 2 K for a field of 50 kOe. In DyRhGe and ErRhGe [Fig. 8(b & c)], MR is negative (-7 % for DyRhGe and -8% for ErRhGe at 50 kOe) near $T_N$. It first increases and becomes positive with field up to the critical field $H_{C1}$. Further increase in field causes the magnitude of MR to decrease. A similar behavior has also been observed in HoRhGe [8]. Thus in compounds with R= Ho, Dy, Er the change in sign is due to the metamagnetic transition (i.e. due to change of magnetic state on the application of field). In the case of ErRhGe, the magnitude of critical field is small as compared to the other compounds and hence when field crosses the critical value, it starts decreasing and becomes negative for further increase in field. The scenario in TbRhGe is somewhat different from the other compounds of this series. The MR is positive below $T_N$, and there is no change in slope beyond the critical field value. It is very interesting that MR linearly increases with field and shows non-saturating behavior even at the highest field. MR in the case of TmRhGe was not calculated due to pulling out of the sample from the resistivity puck on the application of field.

The electronic structure of RRhGe compounds was calculated within the LSDA+$U$ method. The calculated total and partial densities of states (DOS) of RRhGe compounds are shown in Fig. 9. The antiferromagnetic ordering in the rare-earth sublattice results in two classes of R, Rh, and Ge ions. The rare-earth 4f states (of the first class R1) are well defined as sharp peaks from -9 till -6 eV below the Fermi energy ($E_F$) corresponding to the filled 4f states for one spin projection. In the opposite spin projection the partially filled states are located from -6 till -4 eV in different R and the empty states are just above $E_F$. The separation of the filled and empty 4f states is provided by the $U$ correction for the 4f shell in the LSDA+$U$ calculations. The 4f states of the other R2 class have opposite spin projections. The 4d Rh states are non-spin polarized and almost degenerate for Rh1 and Rh2 classes, their DOS is shown as a broad band mostly below $E_F$ down to -6 eV. The other electronic states, including rare-earth 5d and germanium 4p states, are not shown in Fig. 10 but extend over the whole energy range and mix with other states.

In the LSDA+U calculations an antiferromagnetic ordering of the R magnetic moments was obtained as the ground state. The Rh ions were found almost nonmagnetic. To estimate the effective magnetic moment ($\mu_{eff\ LSDA+U}$) of the rare-earth ions from the calculations, one needs to account for orbital moment, due to the neglected spin-orbit coupling. As in work [19], we take into account $L$ = 3 (Tb), 5 (Dy and Tm), 6 (Er), and for the corresponding Lande factor $g$, then the effective magnetic moment of R can be estimated as $g\sqrt{J(J+1)}$, see $\mu_{eff\ LSDA+U}$ values in Table 1. In all compounds the estimated effective magnetic moment values are close to the experimental ones and the calculated free ion moments. In additional LSDA+$U$ calculations with the ferromagnetic and other antiferromagnetic orderings of the R ions moments the total energy of the ferromagnetic solution was found to be higher than that of the antiferromagnetic one in all compounds: by 32 meV in TbRhGe, 26 meV in DyRhGe, 12 meV in ErRhGe, and 15 meV in TmRhGe. Also different types of antiferromagnetic R-ions orderings are close in total energy but still lower than the ferromagnetic one. These values indicate that the antiferromagnetic ordering in ErRhGe, and TmRhGe is slightly weaker than in TbRhGe and DyRhGe, but in general these small energy differences show the possibility to develop modulated structures and magnetic transitions in these intermetallics.

## 4. Summary and Conclusions

All the studied compounds of RRhGe series crystallize in TiNiSi type orthorhombic crystal structure and show antiferromagnetic ordering at low temperatures. Compounds with Tb, Dy and Er show spin reorientation below ordering temperature, as revealed by the heat capacity and magnetocaloric data. The magnetocaloric effect calculated using M-H-T was found to be comparable to that of many well know materials. Field dependence of MR shows anomalous behavior. Below the ordering temperature, the MR in studied compounds is positive, while it is negative above the ordering temperature. In the electronic structure LSDA+$U$ calculations the antiferromagnetic state was obtained as the ground state for all compounds in agreement with experimental data, but the closeness of different antiferro- and ferro-type orderings in total energy promotes the possibility to develop modulated structures and magnetic transitions in these compounds. Therefore, the theoretical results shed some light on to the expected canting, spin reorientation and non collinear magnetic states in many of the members of this series.

## Acknowledgments

SG thanks CSIR, New Delhi for granting senior research fellowship. Authors thank to D. Buddhikot for his assistance in resistivity measurements. The theoretical calculations of the electronic structure were supported by the grant of the Russian Science Foundation (project no. 14-22-00004).

**Figures:**

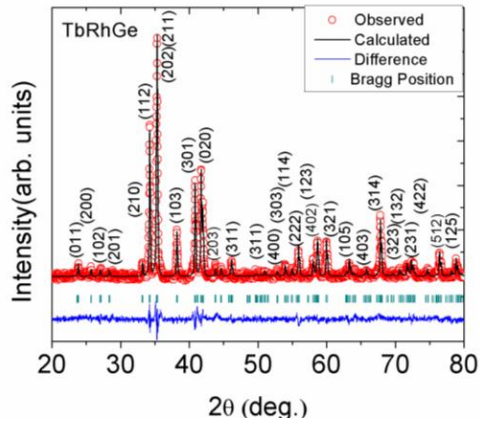

Fig. 1. Powder XRD pattern taken at room temperature along with the Rietveld refinement for TbRhGe. The difference between the observed and the calculated pattern is also shown.

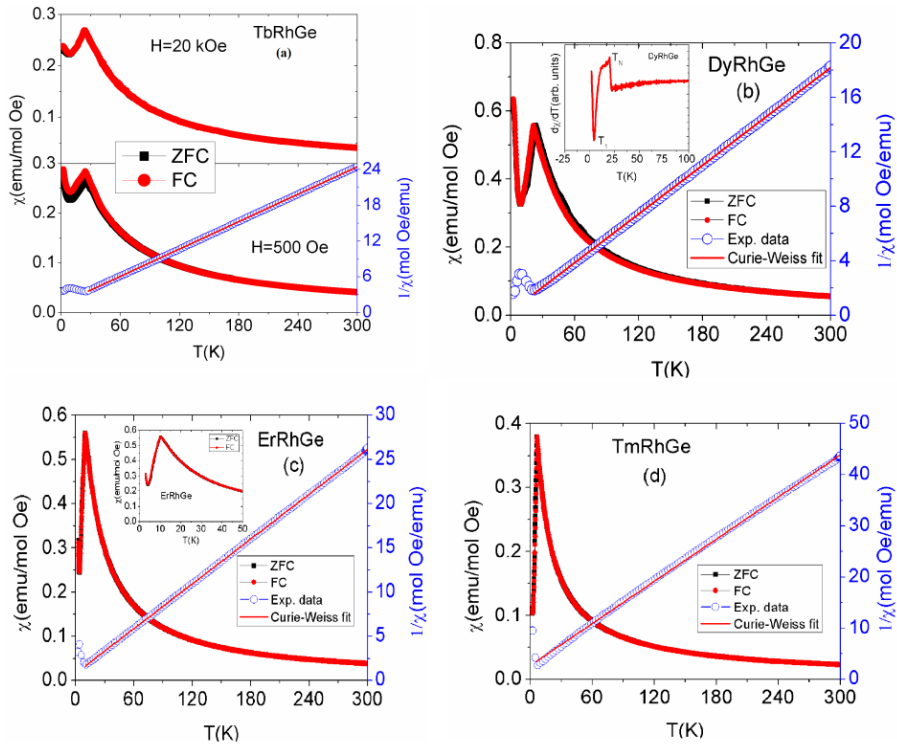

Fig. 2. The temperature dependence of magnetic susceptibility (left-hand scale) and inverse susceptibility along with the Curie-Weiss fit to it (right-hand scale). Insets: (b) temperature derivative of susceptibility for DyRhGe, and (c) low temperature susceptibility plot for ErRhGe.

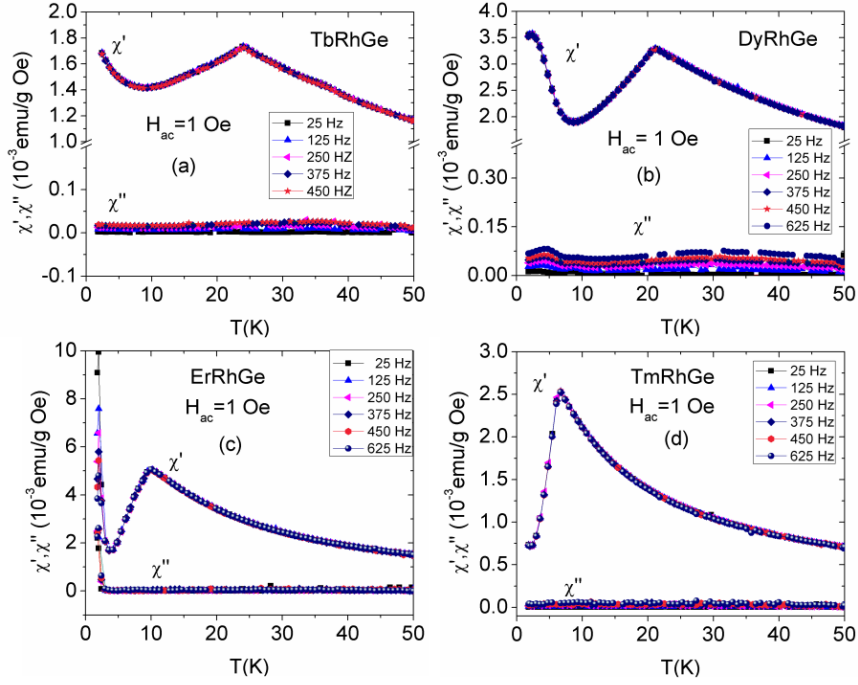

Fig. 3. Temperature dependence of ac susceptibility for the RRhGe compounds at constant ac field ($H_{ac}$=1Oe) and various frequencies (f=25-625 Hz).

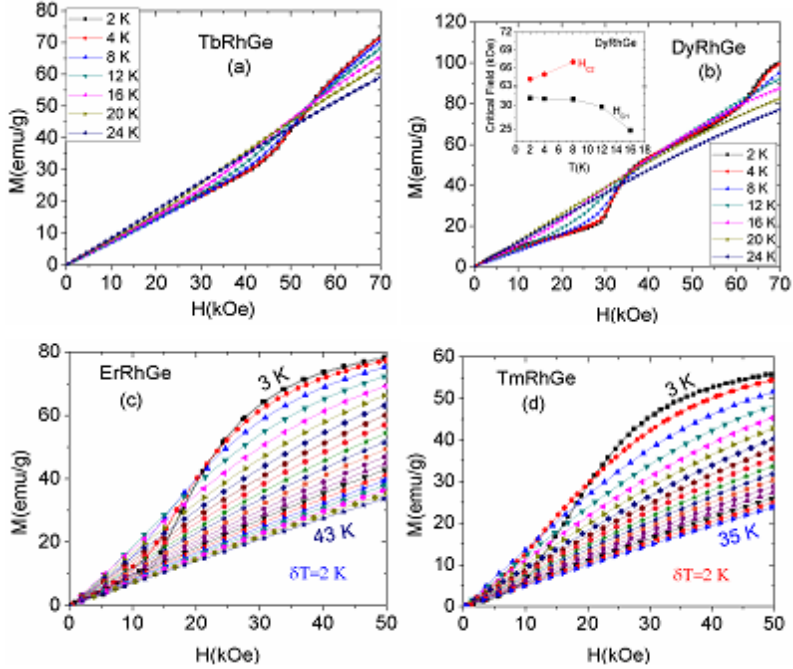

Fig. 4. Magnetization isotherms for the RRhGe compounds obtained at selected temperatures and field up to 70 kOe. Inset in (b) shows the variation of critical fields with temperature in DyRhGe.

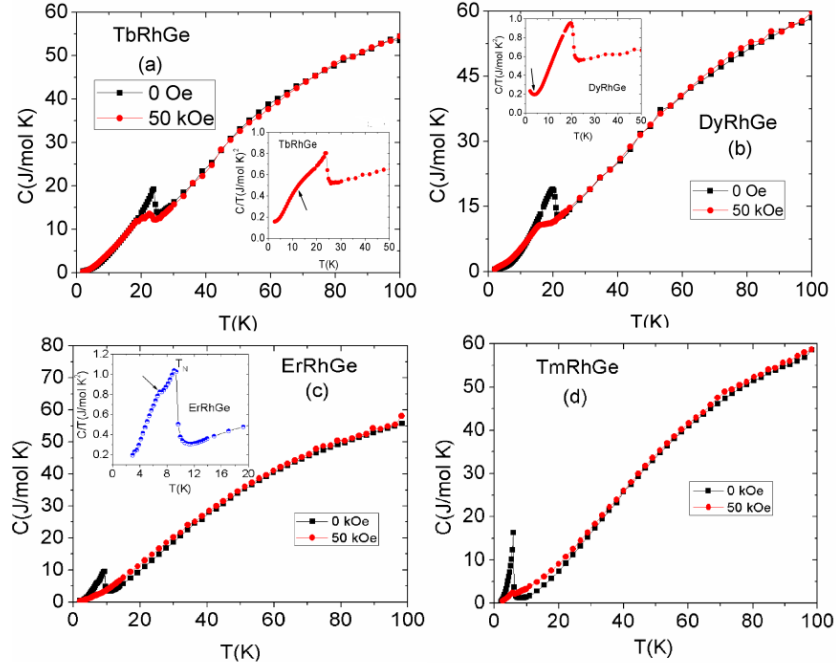

Fig. 5. Temperature dependence of heat capacity of RRhGe compounds in zero and 50 kOe fields. The insets show an expanded plot of zero field heat capacity at low temperatures.

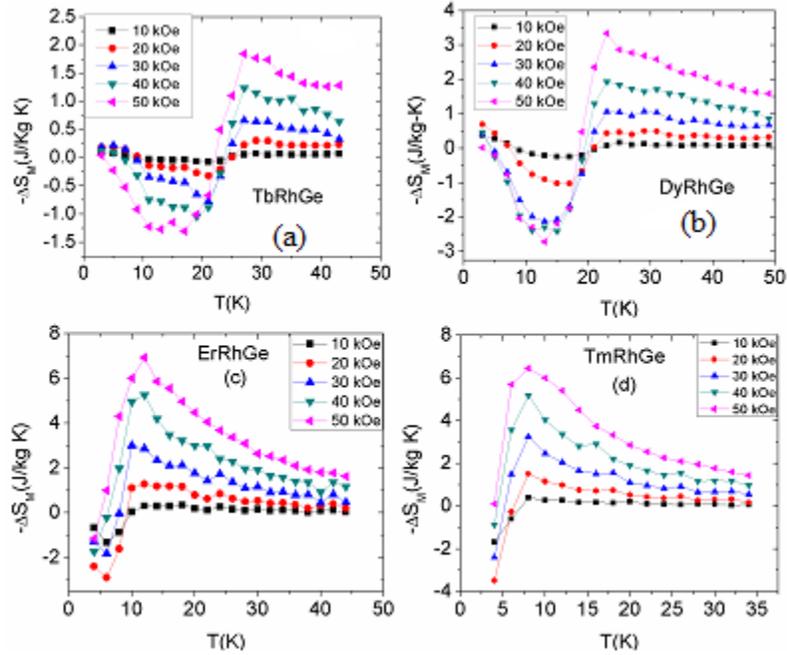

Fig. 6. Temperature dependence of $-\Delta S_M$ at different fields calculated from M-H-T data for the RRhGe compounds.

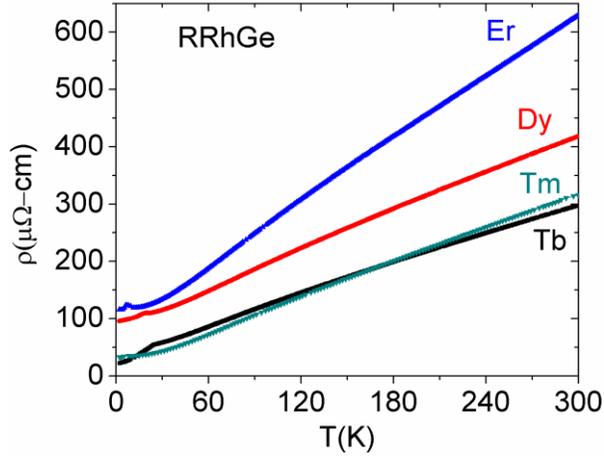

Fig. 7. Temperature dependence of electrical resistivity of RRhGe (R=Tb, Dy, Er, Tm) compounds.

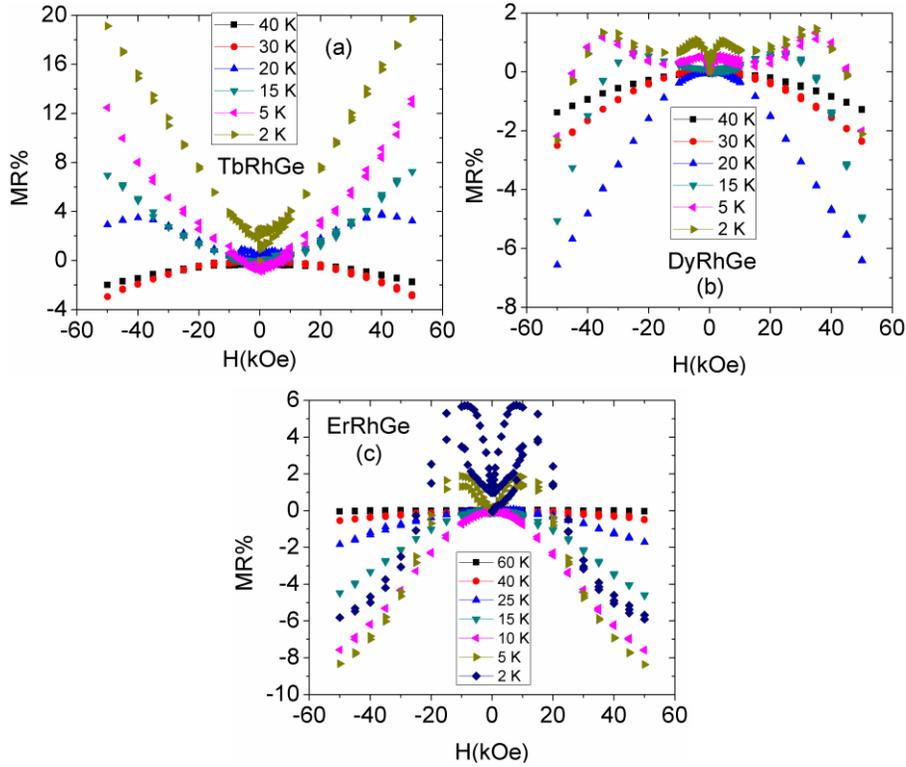

Fig. 8. Field dependence of magnetoresistance at various temperatures for TbRhGe, DyRhGe and ErRhGe compounds.

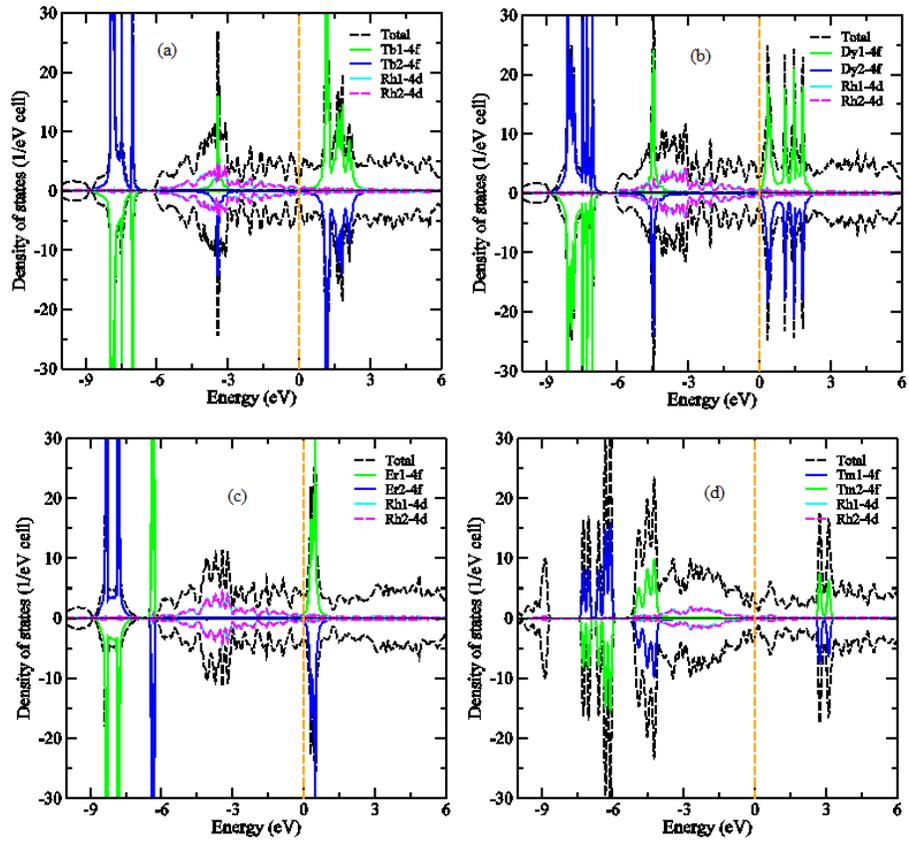

Fig. 9. Total and partial densities of states in TbRhGe (a), DyRhGe (b), ErRhGe (c), and TmRhGe (d) calculated within the LSDA+$U$ method. The Fermi level corresponds to zero.